\documentclass{article}
 \pdfoutput=1 

\usepackage{arxiv}

\usepackage[utf8]{inputenc} 
\usepackage[T1]{fontenc}    
\usepackage{hyperref}       
\usepackage{url}            
\usepackage{booktabs}       
\usepackage{amsfonts}       
\usepackage{nicefrac}       
\usepackage{microtype}      
\usepackage{lipsum}
\usepackage{amsmath}

\usepackage[version=3]{mhchem} 
\usepackage{hyperref}
\usepackage[square,sort,comma,numbers]{natbib} 
\usepackage{graphicx} 

\title{Control of spintronic and electronic properties of bimetallic and vacancy-ordered vanadium carbide MXenes via surface functionalization}

\author{
  Shuo Li \\
  Department of Physical and Macromolecular Chemistry, Faculty of Science \\ Charles University in Prague, 128 43 Prague 2, Czech Republic\\
  \texttt{lishu@natur.cuni.cz} \\
  \And
  Junjie He \\
  Bremen Center for Computational Materials Science \\
  University of Bremen, Am Fallturm 1, 28359 Bremen, Germany
  \And
  Petr Nachtigall  \\
  Department of Physical and Macromolecular Chemistry, Faculty of Science \\ Charles University in Prague, 128 43 Prague 2, Czech Republic\\
  \And
  Luk\'a\v s Grajciar \\
  Department of Physical and Macromolecular Chemistry, Faculty of Science \\ Charles University in Prague, 128 43 Prague 2, Czech Republic\\
  \And
  Federico Brivio $^{\ast}$ \\
  Department of Physical and Macromolecular Chemistry, Faculty of Science \\ Charles University in Prague, 128 43 Prague 2, Czech Republic\\
  \texttt{briviof@natur.cuni.cz} \\
}

\begin{document}
\maketitle

\begin{abstract}
MXenes are 2D transition metal carbides with high potential for overcoming limitations of conventional two-dimensional electronics. In this context, various MXenes have shown magnetic properties suitable for applications in spintronics, yet the number of MXenes reported so far is far smaller than their parental  MAX phases. Therefore, we have studied the structural, electronic and magnetic properties of bimetallic and vacancy-ordered MXenes derived from a new \ce{(V_2/3Zr_1/3)2AlC} MAX phase to assess whether MXene exfoliation would return stable magnetic materials. In particular, we have investigated the properties of pristine and surface-functionalized \ce{(V_{2/3}Zr_{1/3})2CX2} bimetallic and \ce{(V_{2/3}\Box_{1/3})2CX2} vacancy-ordered MXenes with X = O, F and OH. Our density functional theory (DFT) calculations showed that modifying the MXene stoichiometry and/or MXene surface functionalization changes MXene properties. After testing all possible combinations of metallic motifs and functionalization, we identified \ce{(V_{2/3}Zr_{1/3})2CX2}, \ce{(V_{2/3}\Box_{1/3})2CF2} and \ce{(V_{2/3}\Box_{1/3})2C(OH)2} as stable structures. Among them, \ce{(V_{2/3}Zr_{1/3})2CO2} MXene is predicted to be an FM intrinsic half-semiconductor with a remarkably high Curie temperature (\ce{T_C}) of 270 K. The \ce{(V_{2/3}Zr_{1/3})2C(OH)2} MXene exhibits a rather low work function (WF) (1.37 eV) and is thus a promising candidate for ultra-low work function electron emitters. Conversely, the \ce{(V_{2/3}\Box_{1/3})2CF2} MXene has a rather high WF and hence can be used as a hole injector for Schottky-barrier-free contact applications. Overall, our proof-of-concept study shows that theoretical predictions of MXene exfoliation and properties support further experimental research towards developing spintronics devices.
\end{abstract}

\keywords{MXene \and functionalization \and Spintronics}

\section{Introduction}
2D materials are currently the subject of intense experimental and theoretical research for their extremely high aspect ratio, which accounts for their remarkable electronic and magnetic properties. Since graphene\cite{novoselov2004electric} was first isolated, several other ultrathin 2D materials with the same honeycomb lattice have been described, such as boron nitride (h-BN)\cite{lin2009soluble}, silicene\cite{lalmi2010epitaxial}, transition metal dichalcogenides\cite{chhowalla2013chemistry,chhowalla2015two}, monolayer black phosphorus\cite{liu2015semiconducting} and MXenes\cite{naguib201425th}, among others\cite{zhang2015ultrathin}. MXenes are particularly appealing as candidates for spintronics, i.e., the manipulation of electronic spin for logic devices\cite{vzutic2004spintronics}. However, the magnetic response required for spintronics applications is relatively rare in 2D materials\cite{fert2017magnetic}. While magnetism can be introduced in various ways, e.g., via defects or dopants engineering or with external electric fields, our ability to control operational parameters for practical devices remains limited \cite{neto2009electronic}. For example, although phases with high spin-polarization can be achieved in transition metal dichalcogenides, controlling the distribution of dopants and defects is highly difficult \cite{lin20162d,torun2015stable,si2015half}. Accordingly, the design of new intrinsic 2D magnetic materials (which would not require such complicated engineering) would greatly improve their potential for spintronics applications. Several such new classes of compounds have emerged in recent years\cite{gong2017discovery,huang2017layer}, including: half-metals\cite{si2015half,gao2016monolayer,he2016new}, spin gapless semiconductors\cite{ouardi2013realization,wang2008proposal}, bipolar magnetic semiconductors\cite{li2012bipolar} and half-semiconductors\cite{zhang2015robust, liu2016exfoliating,kan2014ferromagnetism}. Among them, half-semiconductors and half-metals have shown the most promising results regarding spin generation, injection, storage and detection \cite{li2016first,zhong2017d}, thus paving the way forward towards a new type of computer components. 

MXenes are 2D transition metal carbides (or nitrides) with the general formula \ce{M2C} (where \ce{M} is a transitional metal) derived from the corresponding MAX phase structures by exfoliation upon chemical etching in aqueous hydrofluoric acid at room temperature\cite{naguib2011two,naguib2012two}. As a result of their synthesis, MXene surfaces are typically functionalized with surface-terminating groups such as \ce{O}, \ce{F} or \ce{OH}\cite{naguib201425th}. Considering their properties, MXenes have been proposed as materials suitable for applications such as transparent conductive films, electromagnetic interference absorption and shielding devices, electrocatalysts, lithium-ion batteries cathods and supercapacitors\cite{ng2017recent,lei2015recent,he2019cr}. In particular, many MXenes have shown potential for spintronics since they are often half semiconductors or half metals\cite{si2015half,gao2016monolayer,he2016new}. In the first case, the conduction band minimum (CBM) and the valence band maximum (VBM) are both fully polarized and both spin channels show a band gap. Conversely, only one channel shows a band gap, while the other is metallic in half metals. Several MXenes have also been predicted to be either half semiconductors or half metals, albeit with limited experimental evidence\cite{si2015half,gao2016monolayer, he2016new,ingason2016magnetic}. Moreover, the electronic structures of MXenes can be fine-tuned by surface modification\cite{liu2016schottky,khazaei2015oh}. Therefore, MXenes are candidates for other applications\cite{khazaei2016nearly, min2006tunable}, such as emitter cathodes in light emitting diodes and field effect transistors\cite{ando2012dependence,freeouf1981schottky}.

The development of MAX phases with bimetallic composition, in particular \ce{(Mo_2/3Sc_1/3)2AlC}\cite{tao2017two,thore2017investigation, khazaei2018electronic} and \ce{(V_2/3Zr_1/3)2AlC}\cite{dahlqvist2017prediction}, has recently opened new MXene research avenues. These bimetallic MXenes (often denoted metal-doped) with functionalized surfaces can be exfoliated by precisely controlling the thermodynamic conditions \cite{tao2017two, thore2017investigation}. In addition, a vacancy-ordered MXene \ce{(Mo_2/3$\Box$_1/3)2C}  (where \ce{$\Box$} denotes the missing transition metal atom) has been reported. Using first-principles calculations, H. Lind \textit{et al.}\cite{lind2017investigation} have computationally shown how introducing different surface functional groups can be used to tune the band gap of \ce{(Mo_2/3$\Box$_1/3)2C}. In this context, we computationally investigated the MAX phase \ce{(V_2/3Zr_1/3)2AlC} and the properties of the corresponding bimetallic \ce{(V_2/3Zr_1/3)2C} MXene and related vacancy-ordered \ce{(V_2/3$\Box$_1/3)2C} MXene. Our aim is to determine whether suitable surface functionalization affects the electronic and magnetic properties of both \ce{(V_2/3Zr_1/3)2CX2} and \ce{(V_2/3\Box_1/3)_2CX2} (X = O, F and OH) MXenes based on first-principles calculations.

\section{Methods}
Density functional theory (DFT) was used to identify structures and electronic and magnetic properties of MXenes along with the \emph{ab-initio} molecular dynamics (AIMD) and Monte Carlo methods for calculating  the kinetic stability and Curie temperature of pristine and functionalized MXenes.
DFT calculations were performed using the Vienna \textit{ab-initio} simulation package (VASP)\cite{kresse1993g,kresse1999g} based on the PAW method. The wave-function was converged to a threshold of $10^{-5}$ $eV$ with a plane-wave energy cut-off set to 500 $eV$ and a k-point mesh of $15 \times 15 \times 1$ following the Monkhorst Pack method for 2D structures. The geometry of all  structures were optimized to converge the interatomic forces below a threshold of 0.01 $eV$/{\AA} using the generalized gradient approximation (GGA) PBE\cite{perdew1996generalized} exchange-correlation functional; the dispersion forces have been described using the DFT-D3 method\cite{grimme2010consistent}. The MXenes unit cell was obtained from the corresponding MAX phase bulk structure, by cutting the \ce{M2C} slab perpendicular to the $<001>$ direction and adding a vacuum region of 15 {\AA}.
To improve the generally underestimated band gaps by GGA methods\cite{he2016new}, the band structure and  projected density of states (PDOS) were obtained with the hybrid HSE06\cite{heyd2003hybrid} functional. The work function (WF) was derived from the energy difference between the Fermi level and the vacuum level\cite{khazaei2015oh}.

Most MXenes have the high-symmetry $P\overline{3}1m$ space group\cite{he2019cr, si2015half,frey2019surface}. In our case, the symmetry of bimetallic and vacancy-ordered MXenes is lower than that of the \ce{V2C} MXene due to the presence of Zr atoms (or vacancy). To accurately assess the band structure and phonon dispersion of MXenes, we expanded the definition of symmetry points of the hexagonal Brillouin zone (see Figure S1), thus breaking the degeneracy of M and K points and naming the now inequivalent points as M1, K1, M2, K2, M3, K3, as previously reported for \ce{(Mo_2/3\Box_1/3)2CX2}\cite{lind2017investigation}. The energy difference at symmetry points is significant and therefore cannot be disregarded. All band structures are in the electronic supplementary information.

The vibrational properties within the harmonic approximation are entirely defined within the dynamical matrix (Hessian matrix) calculated at the density functional perturbation theory (DFPT)\cite{baroni2001phonons} level as implemented in VASP. The post-processing and analysis has thus been performed using the software PhonoPy\cite{togo2015first}. The convergence standards have been increased to $10^{-7}$ $eV$ and to $10^{-6}$ $eV$/{\AA} for the wave-function and interatomic forces, respectively. The other parameters have been kept consistent with the geometry optimization.

To evaluate the stability of functionalized \ce{(V_2/3Zr_1/3)2CX2} MXenes, the formation energy (\ce{E_{form}}) of the unit cell is calculated as:
\begin{equation}
E_{form} =\frac{(E[(V_{2/3}Zr_{1/3})_2CX_2] - E[(V_{2/3}Zr_{1/3})_2C]) - E[X_g])}{3}
\label{eq:form_en}
\end{equation}
where \ce{E[(V_2/3Zr_1/3)2C]} and \ce{E[(V_2/3Zr_1/3)2CX2]} stand for the total energies of pristine \ce{(V_2/3Zr_1/3)2C} and functionalized \ce{(V_2/3Zr_1/3)2CX2}, respectively. \ce{E[X_g]} is the energy of \ce{O2}, \ce{F2} and \ce{H2O} molecules in the gas phase and \ce{E[OH]} = \ce{E[H2O]} - $\frac{1}{2}$\ce{E[H2]}.

For a quick assessment of structural stability, we considered a set of \textit{ab-initio} molecular dynamics (AIMD), as implemented in VASP. These simulations have been completed using the Nos\'e algorithm\cite{nose1984unified} in the NVT ensemble at room temperature (300 K) for  9 ps.

We used a scalar, collinear magnetic model, and we considered the ferromagnetic (FM) configuration and the three possible antiferromagnetic (AFM) states to calculate the preferred magnetic ground state structures of \ce{(V_2/3Zr_1/3)2CX2} system (Figure S2). To comprehensively assess the magnetic behavior of these MXenes, we have constructed an Ising model (see Figure S3) using exchange coupling parameters derived from DFT simulations. This allowed us to calculate the Curie temperature by Monte Carlo simulations (see Section 3.2.3) performed with the open-source software ALPS\cite{albuquerque2007alps}.

\section{Result and discussion}
\subsection{Structural analysis}
\begin{figure}[h]  
\centering
  \includegraphics[width=8cm]{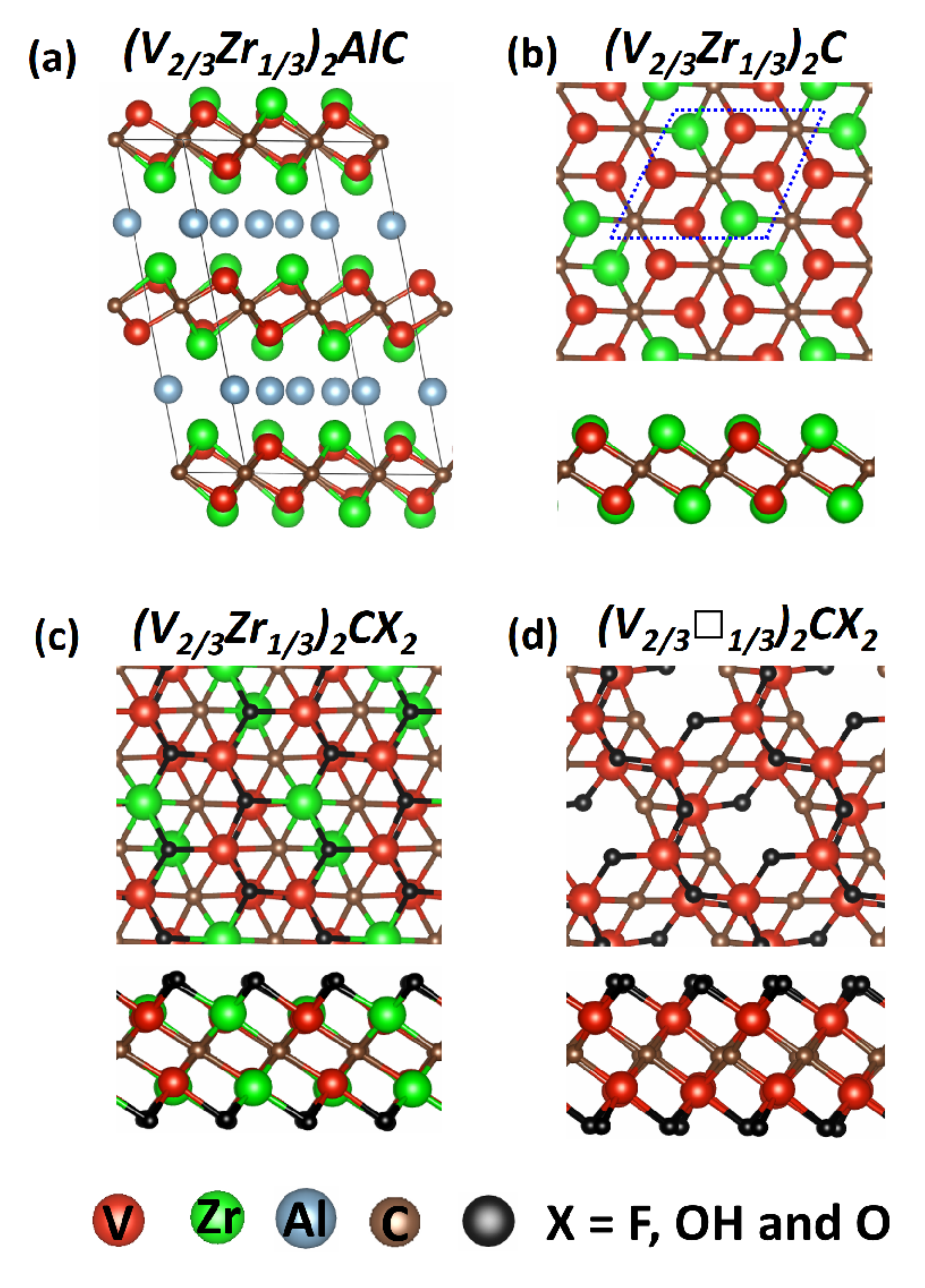} 
  \caption{Structures of parent MAX phase \ce{(V_{2/3}Zr_{1/3})2AlC} (a), pristine bimetallic MXene \ce{(V_{2/3}Zr_{1/3})2C} (b), functionalized bimetallic MXene \ce{(V_{2/3}Zr_{1/3})2CX2} (c) and vacancy-ordered \ce{(V_{2/3}\Box_{1/3})2CX2} (X = F, OH and O) MXene (d). The unit cell of MXenes is marked by a blue dotted line.}
  \label{fig:structure}
\end{figure}

MXene structures derive from corresponding bulk MAX phases, as shown in Figure \ref{fig:structure}. The pristine \ce{(V_2/3Zr_1/3)2C} MXene slab is formed (upon removal of the Al layer) by three hexagonal layers stacked on top of each other, and the layer of C atoms is sandwiched between layers composed of V and Zr atoms (Figure \ref{fig:structure}b). The \ce{Zr} atoms are all aligned along the short diagonal of the \emph{ab}-plane. Functionalized MXenes are obtained by surface termination (F, OH, or O termination), while vacancy-ordered MXenes are obtained by removing Zr atoms. The structural details (e.g., lattice constant, bond lengths, etc.) are outlined in Table S1.

Our calculations suggest that the bimetallic \ce{(V_{2/3}Zr_{1/3})2C} MXene can be obtained from a parent \ce{(V_{2/3}Zr_{1/3})2AlC} MAX phase. First, the calculated exfoliation energy of \ce{(V_2/3Zr_1/3)2CAl} is very similar to that of \ce{V2AlC} \cite{guo2017new} (2.59 \ce{J/m^2} versus 2.53 \ce{J/m^2}, respectively, as shown in Figure S4). Second, the absence of soft modes at the $\Gamma$-point, as shown by the dispersion of the vibrational modes, ensures the dynamical stability of the structure, which is further supported by AIMD calculations performed at 300 K for \ce{(V_2/3Zr_1/3)2C} (Figure S5). All these results indicate that bimetallic MXene is kinetically stable and can be obtained experimentally, similarly to \ce{V2C} \cite{liu2017preparation}. 

The stability and other properties of MXenes depend on surface functionalization, which in turn depends on the exfoliation process. Experimental investigation of related \ce{V2C} MXenes\cite{harris2015direct} has shown that different conditions (i.e. solvent, \ce{O2} partial pressure, etc) during the exfoliation process, lead to different surface decorations\cite{naguib201425th,naguib2011two, khazaei2013novel}. In analogy to the study by Harris \textit{et al.}, we selected the following surface functional groups: \ce{O}, \ce{F} and \ce{OH}. The functional groups are positioned above the \emph{hollow} site formed by three neighbouring \ce{C} mirroring the positions of the metals on the opposite layer. Such structures are systematically more stable than structures with functional groups sitting on top of the \ce{C} atoms (Figure S6).

The formation energies of \ce{(V_2/3Zr_1/3)2CX2}-functionalized MXenes calculated from Equation \ref{eq:form_en} are large, -5.21, -7.55 and -4.48 $eV$ for \ce{X} = \ce{F}, \ce{OH} and \ce{O}, respectively, thus highlighting the formation of strong chemical bonds on MXene surfaces (V or Zr atoms). This is also supported by AIMD simulation at 300K (Figure S7). However, when removing \ce{Zr}, the \ce{(V_2/3\Box_1/3)2C} and \ce{(V_2/3\Box_1/3)2CO2}) structures are subjected to large interatomic forces and heavily distorted during just a short AIMD simulation. We also found that only \ce{(V_2/3\Box_1/3)2CF2} and \ce{(V_2/3\Box_1/3)2C(OH)2} are kinetically stable, based on both AIMD and phonon spectra calculations (Figure S8). Thus, the interaction between metals and surface functional groups is weaker in vacancy-ordered MXenes than in bimetallic  MXenes.

\subsection{Magnetic and electronic properties}
\subsubsection{Bimetallic MXenes}
\begin{figure}[h]
\centering
  \includegraphics[width=8cm]{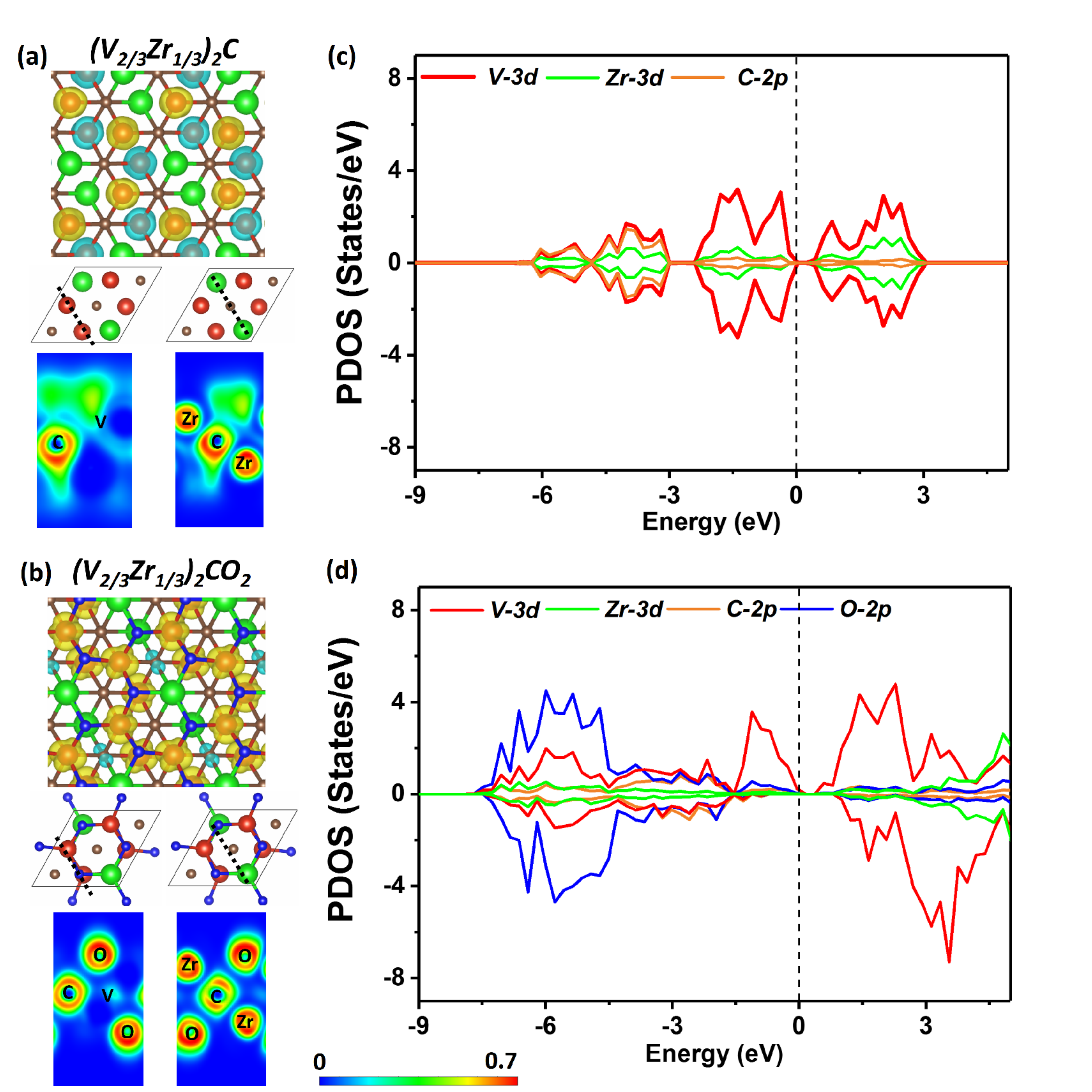} 
  \caption{Spin polarized charge densities and electron localization function (ELF) maps (perpendicular to (001) direction) are shown for \ce{(V_2/3Zr_1/3)2C} (a) and \ce{(V_2/3Zr_1/3)2CO2} (b), where spin up and spin down densities are shown in yellow and light blue, respectively. The units of color scale is "probability". Project density of states (PDOS) of \ce{(V_2/3Zr_1/3)2C} (c) and \ce{(V_2/3Zr_1/3)2CO2} (d). Fermi level (black dotted line) is set to zero.}
  \label{fgr:Fig-2}
\end{figure}

Due to the super-exchange mechanism\cite{kanamori1959superexchange}, \ce{(V_2/3Zr_1/3)2C} exhibits an antiferromagnetic behaviour. This can be seen in the spin-polarized charge densities and in  electron localization functions (ELF) plotted in (Figure \ref{fgr:Fig-2}a).

Electronic Local Function (ELF) maps show that \ce{(V_2/3Zr_1/3)2C} has distinct characteristics of electrons localized on \ce{V} atoms, which leads to the super-exchange mechanism. The band structure of \ce{(V_2/3Zr_1/3)2C} shows a semiconducting character with a band gap of 0.78 $eV$ (Figure S8), and PDOS shown in Figure \ref{fgr:Fig-2}c are in line with ELF. The states near CBM and VBM have main contributions from the \ce{V} $3d$ orbitals, while the contributions from \ce{Zr} $3d$ orbitals and \ce{C} $2p$ orbitals are insignificant.

The magnetic properties of \ce{(V_2/3Zr_1/3)2CO2} are qualitatively different. The presence of surface oxygens changes the magnetic ordering, resulting in FM ground state with  \ce{\Delta E = E_{AFM} - E_{FM}} of 33.59 $meV$/unit cell with respect to the most stable AFM state. The total magnetic moment of \ce{(V_2/3Zr_1/3)2CO2} is 4 $\mu$B per unit cell. $d$-electrons of \ce{V} atoms can induce the spin-polarization of the neighbouring C atoms via double-exchange mechanism \cite{de1960effects}(Figure \ref{fgr:Fig-2}b). The presence of oxygens on the MXene surface also leads to charge transfer towards the surface (as shown by Bader charge analysis\cite{henkelman2006fast}, Table S1), and this affects the position of the \ce{V} $3d$ in the spin-down channel (see PDOS, Figure \ref{fgr:Fig-2}d), which is shifted down in energy and no longer participates in VBM. However, surface oxygens have no effect on the spin-up channel. This is reflected on the band structure, which exhibits distinct half-semiconductor features. The two half-semiconducting gaps are 0.53 and 1.85 $eV$ for spin-up and spin-down channels (Figure S8), respectively. The difference of the band edge energy between the two spin channels (\ce{\Delta E_{CBM} = E_{CBM(down)} - E_{CBM(up)}} and \ce{\Delta E_{VBM} = E_{VBM(down)} - E_{VBM(up)}}) shows the typical half-semiconducting character (See Table S2 for complete set of characteristics). 

Both \ce{(V_2/3Zr_1/3)2CF2} and \ce{(V_2/3Zr_1/3)2C(OH)2} MXenes exhibits similar magnetic properties, and only the results regarding the former are discussed here (see Figure S9 for further details on the latter). The \ce{(V_2/3Zr_1/3)2CF2} MXene exhibits an AFM magnetic character, similarly to pristine \ce{(V_2/3Zr_1/3)2C}. The AFM magnetic configuration is remarkably stable, showing \ce{\Delta E} = -841.08 $meV$. Such analogous value has already been reported for \ce{V2CF2}\cite{hu2014investigations}. The analysis of spin-polarized densities and ELF suggests that  \ce{(V_2/3Zr_1/3)2C(OH)2} MXene shows an analogous behaviour to the pristine MXene, i.e., super-exchange mechanism. The presence of \ce{F} surface atoms widens the band gap due to electron density localization around the halide centre, in analogy to the \ce{O}-terminated surface. However, the FM state is not stabilized in \ce{(V_2/3Zr_1/3)2CF2} since both \ce{V} $3d$ spin channels are equally shifted (Figure \ref{fgr:Fig-3}a and c). 

\subsubsection{Vacancy-ordered MXene}
\begin{figure}
\centering
  \includegraphics[width=8cm]{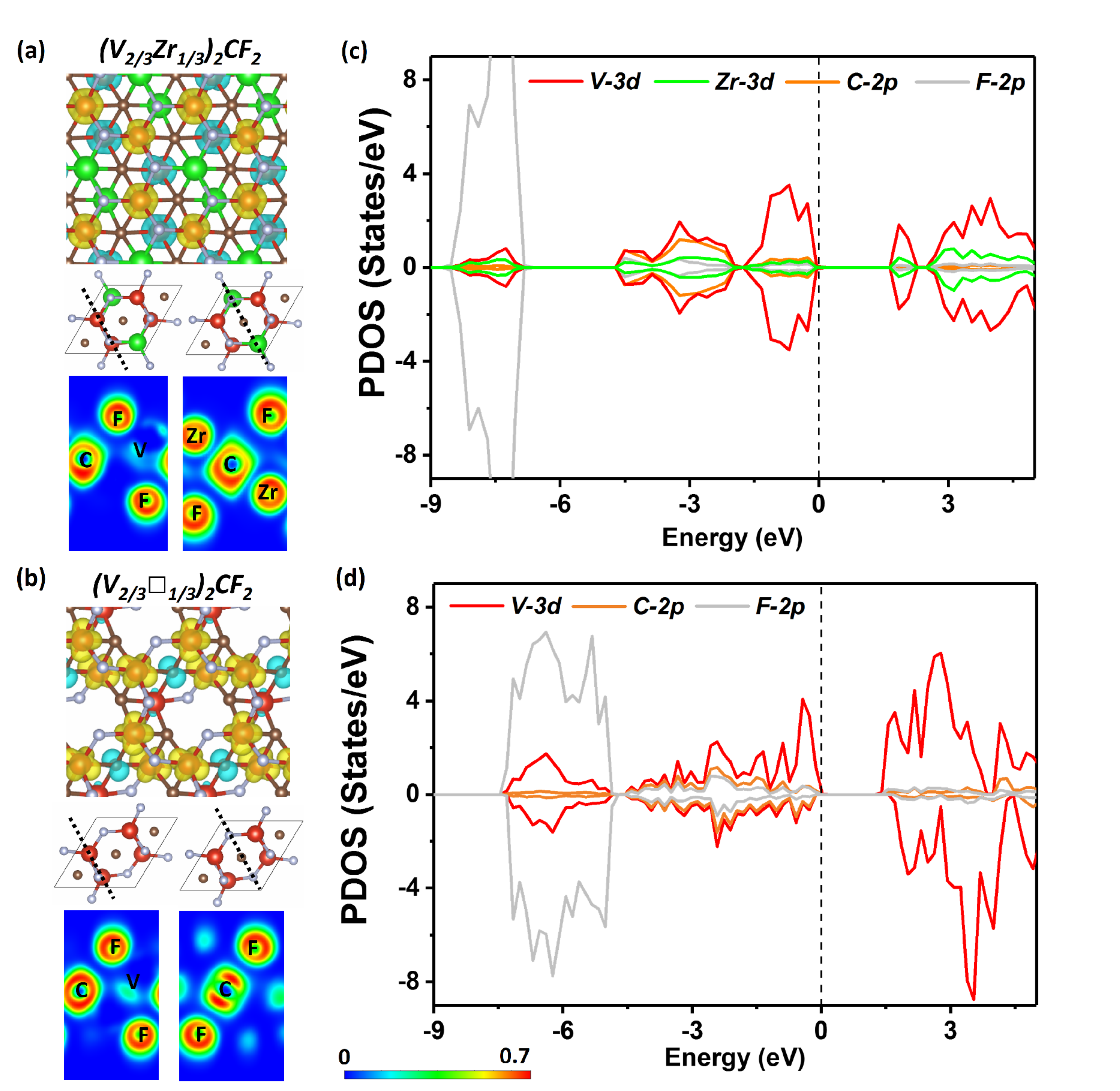} 
  \caption{(a) and (b) The inserted background figure shows spin-polarized charge densities on \ce{(V_2/3Zr_1/3)2CF2} and \ce{(V_2/3\Box_1/3)2CF2}, where spin up and spin down densities are shown in yellow and light blue, respectively. The sections (dot lines) of electron localization function (ELF) maps are perpendicular to (001) direction. The units of color scale is "probability". (e) and (f) The project density of states (PDOS) of \ce{(V_2/3Zr_1/3)2CF2} and \ce{(V_2/3\Box_1/3)2CF2}.  Fermi level is set up to zero with the black dotted line.}
  \label{fgr:Fig-3}
\end{figure}

The pristine \ce{(V_2/3\Box_1/3)2C} MXene is unstable and therefore cannot be directly compared to the pristine bimetallic MXene. Only the \ce{F}-functionalized MXene can be directly compared to the pristine \ce{(V_2/3\Box_1/3)2C} MXene. In contrast to \ce{(V_2/3Zr_1/3)2CF2}, the FM phase of \ce{(V_2/3\Box_1/3)2CF2} is the most stable, with a $\Delta E$ of 22.93 $meV$. The reduced presence of metals induces an overall magnetic moment of 2 $\mu$ B, with the unpaired spins localized on \ce{V} atoms (Figure \ref{fgr:Fig-3}b). 

The FM ordering results from the effect of \ce{V} $d$-electrons, which induce spin-polarization of neighbouring \ce{C} atoms via a double-exchange mechanism, as shown by the ELF map in Figure \ref{fgr:Fig-3}b. The band structure shows a half-semiconducting behaviour (Figure S8), which originates from the spin-polarization of CBM observed in \ce{(V_2/3\Box_1/3)2CF2}. We can also observe a smaller polarization on VBM, due to the slight shift in the PDOS of the V $3d$ orbitals in the spin-down channel. The full set of electronic properties are outlined in Table S2. The polarization of the band edges can be explained by PDOS, wherein, in addition to the \ce{V} $3d$ orbital, \ce{C} and \ce{F} orbitals also significantly contribute to the VBM (\ref{fgr:Fig-3}d). We also considered the \ce{(V_2/3\Box_1/3)2C(OH)2}, which behaves similarly to \ce{(V_2/3\Box_1/3)2CF2}; however, due to the lower electronegativity of \ce{OH}, the effect of spin polarization is weaker and hence $\Delta E$ is smaller.

\subsubsection{Curie temperatures}
\begin{figure}[h]
\centering
  \includegraphics[width=8cm]{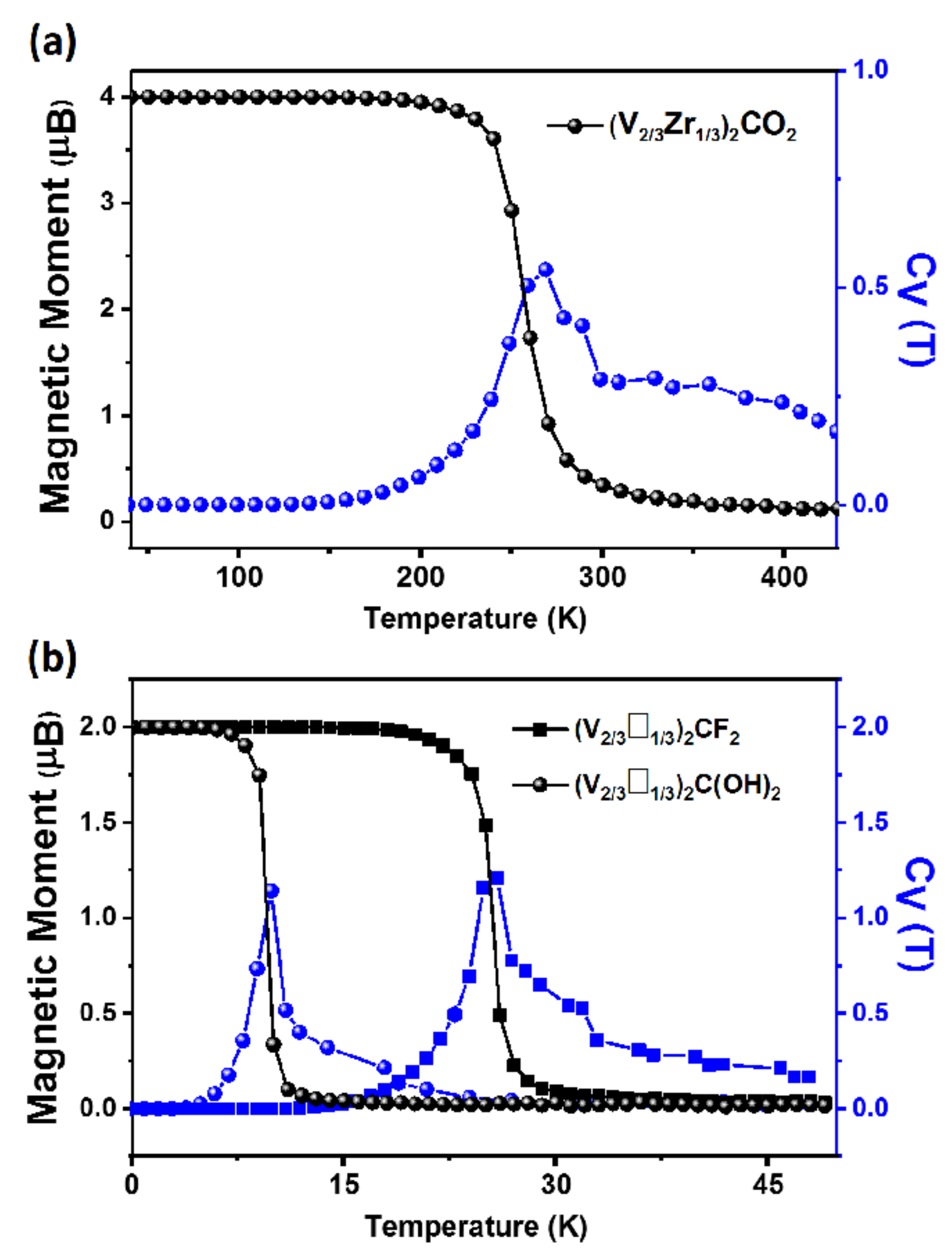} 
  \caption{Variation of the total magnetic momentum (per unit cell) of (a) \ce{(V_2/3Zr_1/3)2CO2} (b) \ce{(V_2/3\Box_1/3)2CF2} and \ce{(V_2/3\Box_1/3)2C(OH)2} with respect to the temperature. Corresponding specific heats \ce{C_V} are shown in blue.
}
  \label{fgr:Fig-4}
\end{figure}

Curie temperatures were calculated for MXene with FM ground state - \ce{(V_2/3Zr_1/3)2CO2}, \ce{(V_2/3\Box_1/3)2CF2} and \ce{(V_2/3\Box_1/3)2C(OH)2}) - using the Monte Carlo simulations. The temperature evolution of magnetic momentum per unit cell is shown in Figure \ref{fgr:Fig-4}. The phase transition between the FM and paramagnetic states for \ce{(V_2/3Zr_1/3)2CO2} occurs around room temperature with an estimated \ce{T_C} of 270 K. This is significantly different from \ce{(V_2/3\Box_1/3)2CF2} and \ce{(V_2/3\Box_1/3)2C(OH)2} MXenes where the transitions occur around 26 and 10 K, respectively. This is due to a weaker magnetic interaction upon the \ce{Zr} atom removal.

\subsubsection{Work function}
\begin{figure}[h]
\centering
  \includegraphics[width=8cm]{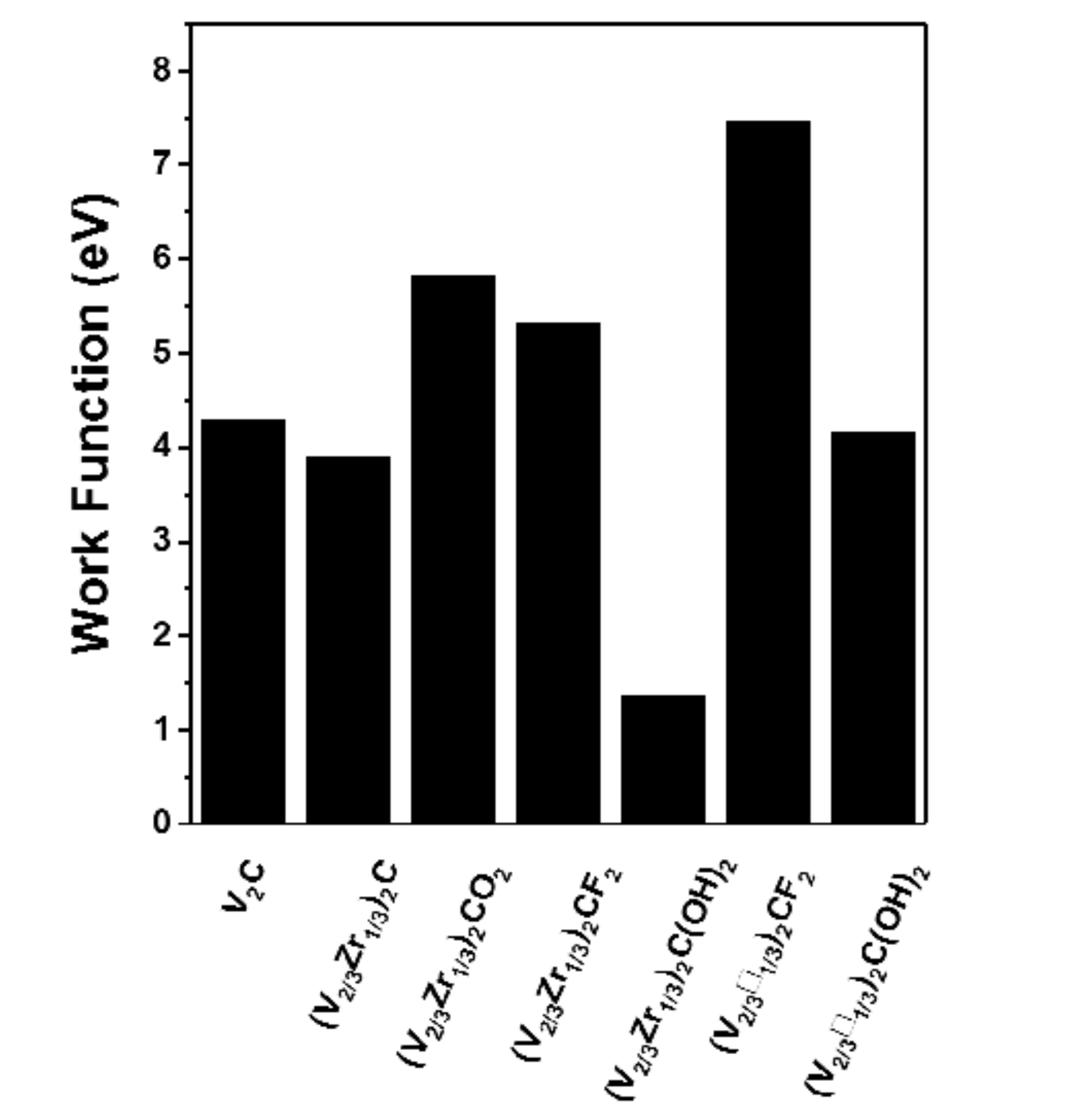} 
  \caption{Work functions of \ce{V2C}, pristine \ce{(V_2/3Zr_1/3)2C}, bimetallic \ce{(V_2/3Zr_1/3)2CX2} and vacancy-ordered \ce{(V_2/3\Box_1/3)2CX2} MXenes}
  \label{fgr:Fig-5}
\end{figure}

The WF of MXenes strongly depends on stoichiometry and surface functionalization (see Figure \ref{fgr:Fig-5}), which makes it possible to obtain materials which can be  used as either electron emitters with low WF or contacts without Schottky barrier and high work-function\cite{liu2016schottky,khazaei2015oh}. The presence of electron-richer Zr in \ce{(V_2/3Zr_1/3)2C} MXene shifts the Fermi level and lower the WF from 4.30 $eV$ with respect to \ce{V2C} to 3.90 $eV$. Surface functionalization affects the electrostatic potential charge near the surfaces\cite{khazaei2016nearly,khazaei2015oh}, thereby increasing the WF of \ce{(V_2/3Zr_1/3)2C} MXene to 5.83 and 5.32 $eV$ when the surface is terminated with \ce{O} or \ce{F}, respectively. A different effect is observed in the \ce{OH} surface termination, where the WF is decreased to 1.37 $eV$ due to the intrinsic dipole of the \ce{OH} group. This value is lower than that previously reported for \ce{Sc2C(OH)2} MXene (1.60 $eV$)\cite{khazaei2015oh}. Therefore, \ce{(V_2/3Zr_1/3)2C(OH)_2} MXene is a candidate for  low-WF emitting cathods. Removing the \ce{Zr} atoms increases the WF  to 7.47 and 4.16 $eV$ for \ce{(V_2/3\Box_1/3)2CF2} and \ce{(V_2/3\Box_1/3)2C(OH)2} MXenes, respectively. Our analysis clearly shows that the WF of various \ce{(V_2/3Zr_1/3)2C} MXenes investigated herein can be tuned within a broad range of values; the \ce{(V_2/3Zr_1/3)2C(OH)2} MXene could be used as an ultra-low work function electron emitter. It's interesting to note that the \ce{(V_2/3\Box_1/3)2CF2} MXene has a higher WF than the Pt metal (which has the highest WF among the elemental metals\cite{liu2016schottky}), it could be used for hole injection in Schottky-barrier-free contact applications.

\section{Conclusions}
Our results show that modifying the stoichiometry and/or surface functionalization of MXenes changes their properties qualitatively. Therefore they are excellent candidates for applications in spintronics because their electric and magnetic properties can be tuned for specific purposes. In this study,
we identified \ce{(V_{2/3}Zr_{1/3})2CX2}, \ce{(V_{2/3}\Box_{1/3})2CF2} and \ce{(V_{2/3}\Box_{1/3})2C(OH)2} as stable candidates. 
Among them, the \ce{(V_2/3Zr_1/3)2CO2} and \ce{(V_{2/3}\Box_{1/3})2CF2} and \ce{(V_{2/3}\Box_{1/3})2C(OH)2} are half-semiconductor materials. 
The predicted Curie temperature for \ce{(V_2/3Zr_1/3)2CO2} (270 K) is higher than that of the experimentally reported 2D \ce{CrI3} crystals (45 K).\cite{huang2017layer} Hence, of the MXenes tested in this study, \ce{(V_2/3Zr_1/3)2CO2} is the best candidate for spintronic applications. 

The functional groups of MXenes can radically change the composition of their  frontier orbitals, thereby affecting their work function. In particular, we found that \ce{(V_2/3Zr_1/3)2C(OH)2} MXene can be used as an ultra-low work function electron emitter and its work function (\ce{1.37} $eV$) is lower than that of \ce{Sc2C(OH)2} MXene  (1.6 $eV$), as reported by Khazaei \emph{et al.}. Conversely, the \ce{(V_{2/3}\Box_{1/3})2CF2} MXene has a rather high WF of 7.47 $eV$, which is thus higher than that of the Pt metal\cite{liu2016schottky}. Thanks to these properties, \ce{(V_{2/3}\Box_{1/3})2CF2} MXenes can be used for holes injection in applications requiring Schottky-barrier-free contacts. Therefore, \ce{(V_2/3Zr_1/3)2C(OH)2} and vacancy-ordered \ce{(V_{2/3}\Box_{1/3})2CF2} MXenes are also promising candidates for electronic devices. 

Overall, the results presented in this study establish a new family of MXenes with intrinsic magnetism, which makes them ideal candidates for both spintronic and electronic applications in the near future.

\section{Acknowledgements}
A support from OP VVV ''Excellent Research Teams'', project CUCAM, is also acknowledged. J. H. acknowledges the financial support provided by the National Natural Science Foundation of China (Grant No. 11804041). S. L. acknowledges the support from GAUK project (Grant No. 792218). Thanks Carlos V. Melo for his precious contribution of writing guidance.

\bibliography{main}
\bibliographystyle{unsrt}

\end{document}